\documentclass[aps,prd,showpacs,floatfix,nofootinbib,twocolumn,10pt]{revtex4}

\usepackage{color,amsmath,amssymb,graphicx,latexsym,subfigure}
\usepackage{threeparttable}

\newcommand{\sv}{\ensuremath{\langle\sigma v\rangle}}

\begin{document}

\title{Reconcile the AMS-02 positron fraction and Fermi-LAT/HESS
total $e^{\pm}$ spectra by the primary electron spectrum
hardening}

\author{Qiang Yuan$^{a,b}$}
\author{Xiao-Jun Bi$^b$}

\affiliation{
$^a$Key Laboratory of Dark Matter and Space Astronomy, Purple 
Mountain Observatory, Chinese Academy of Sciences, Nanjing
210008, P.R.China \\
$^b$Key Laboratory of Particle Astrophysics,
Institute of High Energy Physics, Chinese Academy of Science,
P.O.Box 918-3, Beijing 100049, P.R.China 
}

\begin{abstract}

The recently reported positron fraction up to $\sim 350$ GeV by
AMS-02 seems to have tension with the total electron/positron
spectra detected by Fermi and HESS, for either pulsar or dark
matter annihilation/decay scenario as the primary positron
sources. In this work we will show that the tension will be
removed by an adjustment of the primary electron spectrum. If
the primary electron spectrum becomes harder above $\sim50$
GeV, similar as the cosmic ray nuclei spectrum, the AMS-02 positron 
fraction and Fermi/HESS data can be well fitted by both the pulsar 
and dark matter models. This result may be suggestive of a 
common origin of the cosmic ray nuclei and the primary electrons. 
Furthermore, this study also implies that the properties 
of the extra sources derived from the fitting to the AMS-02 data 
should depend on the form of background.

\end{abstract}

\date{\today}

%95.35.+d: Dark matter
%96.50.S-: Cosmic rays

\maketitle

\section{Introduction}

The AMS-02 collaboration reported the very precise measurement of the
positron fraction $e^+/e^{\pm}$ with energies up to $350$ GeV recently
\cite{2013PhRvL.110n1102A}. The positron fraction shows a continous
increase up to $\sim 100$ GeV, which is consistent with the previous
PAMELA result \cite{2009Natur.458..607A,2010APh....34....1A} and is
lower than that measured with Fermi-LAT \cite{2012PhRvL.108a1103A}.
A flattening of the positron fraction above $\sim 100$ GeV is revealed
by the AMS-02 data, for the first time. The AMS-02 data implies that
there is excess of positrons above tens of GeV compared with the
standard cosmic ray (CR) background, and the amount of excess positrons
should be less than previously estimated according to the PAMELA data.

Several works appears to explain the AMS-02 data with pulsars
\cite{2013ApJ...772...18L} or dark matter (DM) scenarios
\cite{2013PhRvD..88g6013K,2013JCAP...05..003D,2013JHEP...08..029I}.
A thorough study of the properties of the extra positron sources,
including the astrophysical one like pulsars and the dark matter
(DM) scenario, based on the AMS-02 data and the electron (or
$e^{\pm}$) spectra measured by PAMELA \cite{2011PhRvL.106t1101A},
Fermi-LAT \cite{2009PhRvL.102r1101A,2010PhRvD..82i2004A} and HESS
\cite{2008PhRvL.101z1104A,2009A&A...508..561A}, was given shortly
after the publication of the AMS-02 data \cite{2013arXiv1304.1482Y} 
(Paper I). It was found that there was difficulty to fit the AMS-02 
positron data and the Fermi-LAT/HESS total electron spectra 
simultaneously, either in the pulsar scenario or in the DM scenario. 
The results seem to imply that there might be {\em tension} between 
the AMS-02 data and the Fermi-LAT/HESS data, in the present 
theoretical framework. This conclusion has been confirmed by other
studies\footnote{Also the preliminary data about the total $e^{\pm}$
spectra by AMS-02 show the discrepancy with Fermi-LAT data below
$\sim100$ GeV. For $E\gtrsim100$ GeV the difference between these 
two data sets are smaller \cite{2013ICRC-AMS02}. The discussion 
of the present work may hold given the new AMS-02 $e^{\pm}$ spectra, 
since we study to reconcile the high energy behavior of Fermi-LAT data 
with the AMS-02 positron fraction.} \cite{2013JCAP...11..026J,
2013PhRvD..88b3013C}. 
One possible reason leading to the {\em tension} is the constraint on
the electron injection parameters by the pure electron spectrum by 
PAMELA. If the PAMELA data are not included, the primary electron 
spectrum has larger free space and the AMS-02 data and Fermi data 
can be easier to be fitted simultaneously, as shown in some recent 
works to explain the AMS-02 result \cite{2013ApJ...772...18L,
2013PhRvD..88g6013K,2013JCAP...05..003D,2013JHEP...08..029I}.

Several possibilities to reconcile these two data sets were
discussed in Paper I, including multiple components of the extra
sources and the existence of spectral hardening of the primary
electron spectrum. The idea to introduce a spectrum hardening 
of primary electrons to fit the data was also raised in 
\cite{2014PhLB..728..250F,2013PhRvD..88b3013C}. The spectrum hardening
was stimulated by the observed spectral hardening of the nuclei
spectra in recent years by several collaborations\footnote{Note,
however, the recent reported preliminary data about the proton and
Helium spectra show no significant hardening as observed by PAMELA
\cite{2013ICRC-AMS02}. Whether or not the heavier nuclei
have the hardening still needs to be tested by the future AMS-02
data. Since the present work is based on the electron and positron
data, the AMS-02 results about the nuclei spectra have only a slight
modification of the secondary positron spectrum and do not affect 
the discusion significantly.}
\cite{2007BRASP..71..494P,2010ApJ...714L..89A,
2011Sci...332...69A}. A unified spectral hardening at rigidity
$R\sim 200$ GV (or $E_k\sim200$ GeV/n) was measured precisely by
PAMELA or CREAM \cite{2011Sci...332...69A,2010ApJ...714L..89A}. 
If there is a spectral hardening of the CR nuclei spectra, it is
natural to expect a similar hardening of the primary electron spectrum.

Models to explain the spectral hardening include the multi-component sources
\cite{2006A&A...458....1Z,2011PhRvD..84d3002Y,2012APh....35..449E,
2012MNRAS.421.1209T,2013A&A...555A..48B}, non-linear acceleration
of the particles \cite{2013ApJ...763...47P}, or the propagation
effect \cite{2012ApJ...752...68V,2012ApJ...752L..13T}. In
\cite{2014PhLB..728..250F} the authors pointed out that if there
was a spectral hardening of the primary electron spectrum, there
would be a less steep increase (or decrease) of the positron
fraction above $\sim200$ GeV.

In the work we investigate in detail whether to involve such a
spectral hardening of the primary electron spectrum can help
eliminate the {\em tension} between the AMS-02 data and the
Fermi-LAT/HESS data. We employ the CosRayMC tool developed in
\cite{2012PhRvD..85d3507L} to fit the observational data within
the high dimensional parameter space. The GALPROP 
package\footnote{http://galprop.stanford.edu/}
\cite{1998ApJ...509..212S} to calculate the propagation of the
charged CRs has been embedded in the Markov Chain Monte Carlo 
(MCMC) algorithm, which is well known to be efficient for the 
survey of high-dimensional correlated parameter space 
\cite{2002PhRvD..66j3511L}, in CosRayMC. The
diffusion-reacceleration propagation frame is adopted, and the
major propagation parameters are $D_0|_{R_0=4\,{\rm
GV}}=5.94\times10^{28}$ cm$^2$ s$^{-1}$, $\delta=0.377$,
$v_A=36.4$ km s$^{-1}$ and $z_h=4.04$ kpc \cite{mcmc:prop}. The
goodness of fit, constraints and implication of the model
parameters are discussed.

In the next Section we simply describe the models to fit the data.
The results are presented in Sec. III. In Sec. IV we give the
discussions and conclusions.

\section{Model}

In this section we describe the major aspects of the theoretical
models to reproduce the electron/positron data briefly. The
injection spectra of the primary protons (heavier nuclei are less
important in this study) and electrons are both assumed to be
broken power-law functions with respect to momentum $p$
\begin{eqnarray}
q(p)\propto\begin{cases}
(p/p_{\rm br,1}^{p,e})^{-\gamma_0},&p<p_{\rm br,1}^{p,e}\\
(p/p_{\rm br,1}^{p,e})^{-\gamma_1},&p_{\rm br,1}^{p,e}<p<p_{\rm br,2}^{p,e}\\
(p/p_{\rm br,2}^{p,e})^{-\gamma_2}(p_{\rm br,2}^{p,e}/p_{\rm br,1}^{p,e})^{-\gamma_1},&p>p_{\rm br,2}^{p,e}
\end{cases}
\end{eqnarray}
where $p_{\rm br,1}$ represents the low energy break, $p_{\rm br,2}$
is the high energy break to be responsible for the spectral hardening,
$\gamma_0$, $\gamma_1$ and $\gamma_2$ are the spectral indices in different
momentum ranges. We also employ the log-parabolic function to describe
the spectral hardening of the electrons, i.e., $q(p)\propto (p/p_{\rm br,1}^e)
^{-\gamma_1+\gamma_2\log(p/{\rm MeV})}$ for $p>p_{\rm br,1}^e$. In this
case $p_{\rm br,2}^e$ is not used. The absolute fluxes of protons and
electrons are determined through normalizing the propagated fluxes to
normalization factors $A_p$ and $A_e$.

The background positrons are expected to be produced through the
collision of CR nuclei with the interstellar medium (ISM) during
the propagation. The parameterization of $pp$ collision in
\cite{2006ApJ...647..692K} is employed to calculate the secondary
production of positrons and electrons. Similar as done in Paper I,
we further introduce a free factor $c_{e^+}$ to adjust the absolute
fluxes of the secondary positrons and electrons to fit the data.
Such a factor may represent the uncertainties of the hadronic
interactions, propagation models, the ISM density distributions,
and the nuclear enhancement factor from heavy elements.

In the PAMELA era it was found that the background contribution
are not enough to explain the observed positron fraction and total
$e^{\pm}$ data \cite{2010IJMPD..19.2011F,2012APh....39....2S}.
Therefore the extra sources of $e^{\pm}$ beyond the traditional CR
background are introduced to explain the data.
%A natural assumption is that
%the positrons and electrons are produced with equal amount from
%the extra sources.
We will base on the same theoretical framework to fit the AMS-02
data in the work,  assuming continuously distributed pulsars or
the DM annihilation/decay to be the extra sources of $e^{\pm}$.

The injection spectrum of $e^{\pm}$ from pulsars is assumed to be
power-law with an exponential cutoff
\begin{equation}
q(p)=A_{\rm psr}p^{-\alpha}\exp(-p/p_c),
\end{equation}
where $A_{\rm psr}$ is the normalization factor, $\alpha$ is the spectral
index and $p_c$ is the cutoff momentum. The spectral index $\alpha$ is
limited in the range $1.4$ to $2.2$ according to the $\gamma$-ray
observations of pulsars \cite{1994ApJ...436..229T}. The spatial
distribution of pulsars is taken to be the cylindrically symmetric form
given in \cite{2004IAUS..218..105L}
\begin{equation}
f(R,z)\propto\left(\frac{R}{R_{\odot}}\right)^{2.35}\exp\left[-\frac{5.56
(R-R_{\odot})}{R_{\odot}}\right]\exp\left(-\frac{|z|}{z_s}\right),
\label{source}
\end{equation}
where $R_{\odot}=8.5$ kpc is the distance of the solar location to the
Galactic center, $z_s\approx 0.2$ kpc is the characteristic height
of the Galactic disk.

As for DM scenario (taking annihilation as illustration), we focus on the
leptonic two-body annihilation channels $\mu^+\mu^-$ and $\tau^+\tau^-$,
as implied according to the PAMELA and Fermi-LAT/HESS data of the
electrons/positrons and the antiprotons
\cite{2009NuPhB.813....1C,2009PhRvD..79b3512Y,2010NuPhB.831..178M}.
The positron/electron production function from DM annihilation is
(assumed to be Majorana particles)
\begin{equation}
q(r,p)=\frac{\langle\sigma v\rangle}{2m_{\chi}^2}\frac{dN}{dp}
\times\rho^2(r),
\end{equation}
where $m_{\chi}$ is the mass of DM particle, $\langle\sigma v\rangle$
is the velocity weighted annihilation cross section, $dN/dp$ is the
yield spectrum for one annihilation of a pair of DM particles,
and $\rho(r)$ is the DM density profile. The spatial profile of DM
energy density is taken to be Navarro-Frenk-White (NFW,
\cite{1997ApJ...490..493N}) distribution
\begin{equation}
\rho(r)=\frac{\rho_s}{(r/r_s)(1+r/r_s)^2},
\end{equation}
with parameters $r_s=20$ kpc and $\rho_s=0.26$ GeV cm$^{-3}$.

For low energy particles we further employ a simple force field
approximation to take into account the solar modulation effect
\cite{1968ApJ...154.1011G}. Since the operation period of PAMELA
and AMS-02 is close to the solar minimum, the modulation potential
is required to be smaller than $1$ GV. Note, however, the low energy
part of the positron fraction measured by PAMELA and AMS-02 might not
be reproduced with such single solar modulation model, and more
complicated charge-sign dependent modulation effect is necessary
\cite{2012AdSpR..49.1587D,2013PhRvL.110h1101M}.

\section{Results}

\begin{figure}[!htb]
\includegraphics[width=0.45\textwidth]{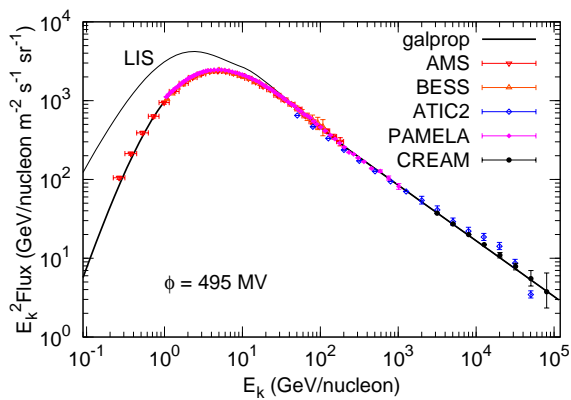}
\caption{Proton spectrum derived through fitting PAMELA and CREAM data.
References of the proton data: AMS \cite{2000PhLB..490...27A},
BESS \cite{2000ApJ...545.1135S}, ATIC2 \cite{2007BRASP..71..494P},
PAMELA \cite{2011Sci...332...69A} and CREAM \cite{2010ApJ...714L..89A}.
\label{fig:proton}}
\end{figure}

We first determine the parameters of the proton injection spectrum
through fitting to the PAMELA \cite{2011Sci...332...69A} and CREAM
\cite{2010ApJ...714L..89A} data. For CREAM data we include $10\%$
systematic uncertainties as discussed in
\cite{2010ApJ...714L..89A}. The high energy break $p_{\rm br,2}^p$
is fixed to be $230$ GeV as suggested by the PAMELA data. The best
fitting parameters of the proton spectrum are: $\gamma_0=1.80$,
$\gamma_1=2.42$, $\gamma_2=2.33$, $p_{\rm br,1}^p=12.3$ GeV, and
solar modulation potential $\phi=495$ MV. The normalization of the
propagated proton flux at $100$ GeV is $A_p=4.55\times10^{-9}$
cm$^{-2}$s$^{-1}$sr$^{-1}$MeV$^{-1}$. Comparison of the best
fitting spectrum of protons with the observational data is shown
in Fig. \ref{fig:proton}. We see very good agreement between the
calculated spectrum and the data. The minimum $\chi^2$ value is
about $24$ for $72$ degree of freedom (dof).

Since the observational period of protons by PAMELA is almost the same
with that of electrons by PAMELA and positrons by AMS-02, we should
expect a common modulation amplitude for these particles (besides the
charge-sign dependent effect). Therefore we employ a prior on the
modulation potential $\phi=500\pm53$ MV comes from the fit of the
proton data.

\begin{figure*}[!htb]
\includegraphics[width=0.45\textwidth]{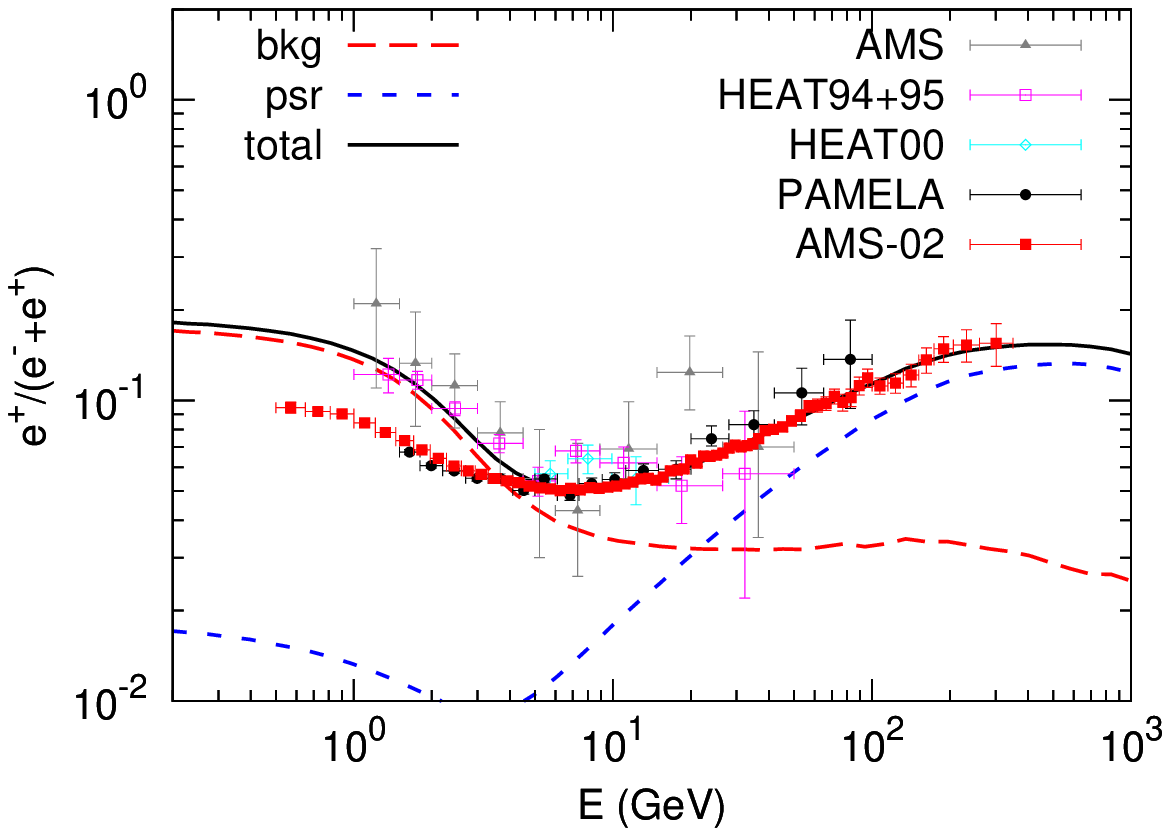}
\includegraphics[width=0.45\textwidth]{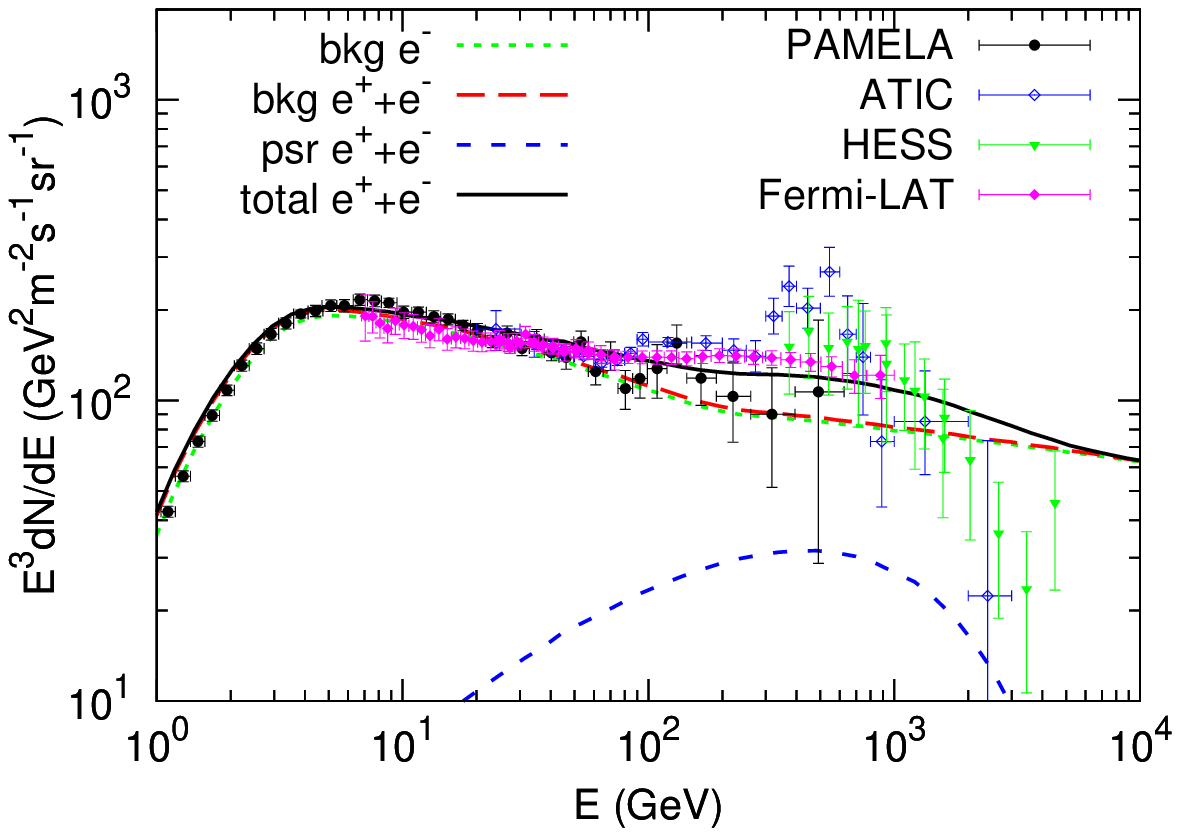}
\caption{The positron fraction (left) and electron spectra (right)
for the background together with a pulsar-like component of the exotic
$e^{\pm}$. The high energy hardening of the primary electron spectrum
is approximated with a broken power-law and the break momentum
$p_{\rm br,2}^e$ is fixed to be $\sim230$ GeV. References of the data:
positron fraction --- AMS01 \cite{2007PhLB..646..145A},
HEAT94+95 \cite{1997ApJ...482L.191B}, HEAT00 \cite{2001ICRC....5.1687C},
PAMELA \cite{2009Natur.458..607A}, AMS-02 \cite{2013PhRvL.110n1102A};
electron --- PAMELA \cite{2011PhRvL.106t1101A},
ATIC \cite{2008Natur.456..362C}, HESS \cite{2008PhRvL.101z1104A,
2009A&A...508..561A}, Fermi-LAT \cite{2010PhRvD..82i2004A}.
\label{fig:psr_break2}}
\end{figure*}

%In the following discussion we primarily consider pulsars as the
%sources of the extra positrons and electrons. The discussion of DM
%models are left in the end of this section.
We first fix the electron second break energy $p_{\rm br,2}^e$ at
$230$ GeV, same as that of protons. The best fitting results of
the positron fraction and electron spectrum are shown in Fig.
\ref{fig:psr_break2}. The fitting parameters and the $\chi^2$
value are presented in Table \ref{table:psr}. Compared with the
case without spectral hardening of the primary electrons (Paper
I), the fitting is indeed improved. The $\chi^2$ value decrease
from $\sim280$ to $235$ with one more parameter. However, the
overall fitting is still not satisfactory. We can see from Fig.
\ref{fig:psr_break2} that when AMS-02 data are well reproduced,
the model expectation is lower than the Fermi data, which is
similar with the findings in Paper I.

\begin{figure*}[!htb]
\includegraphics[width=0.45\textwidth]{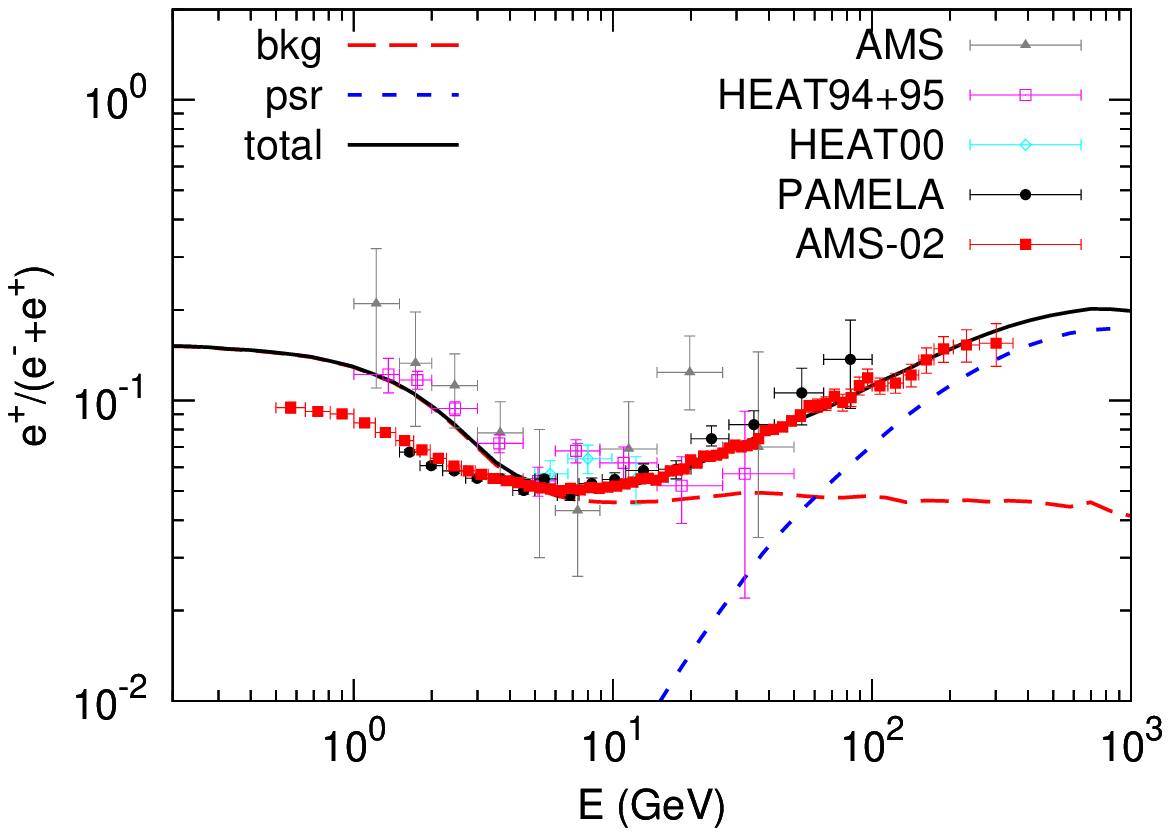}
\includegraphics[width=0.45\textwidth]{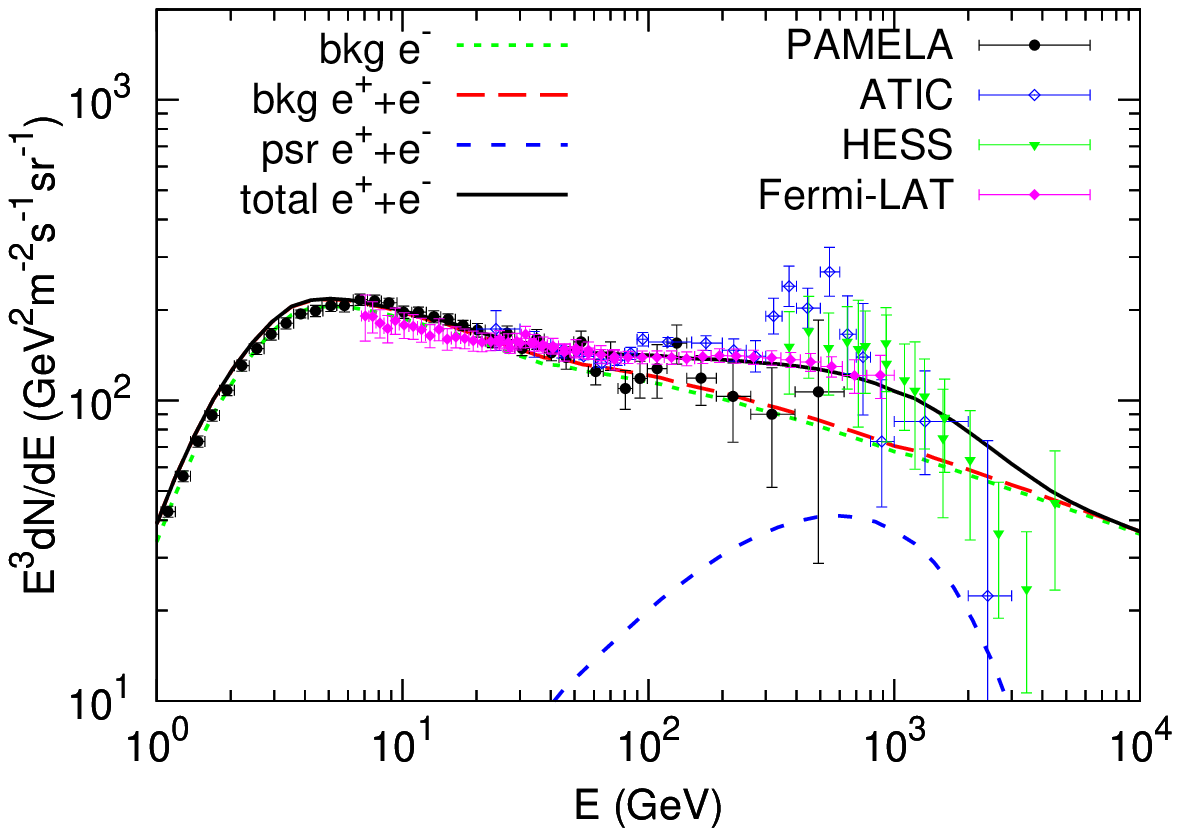}
\caption{Same as Fig. \ref{fig:psr_break2} but the high energy break
of the electrons is relaxed in the fit.
\label{fig:psr_break}}
\end{figure*}

We then relax the break momentum of the electrons and redo the
fit. In this case we find the improvement is significantly, as
shown in Fig. \ref{fig:psr_break}. The parameters are also given
in Table \ref{table:psr}. The minimum $\chi^2$ value over dof is
about $1.0$, which implies a rather good fitting. However, the
break momentum $p_{\rm br,2}^e$ is required to be about $45$ GeV,
which is significantly smaller than that of the nuclei. The
difference of the spectral indices below and above $p_{\rm
br,2}^e$ is about $0.3$. As a comparison, such a value is measured
to be $\sim0.2$ for protons and $\sim0.3$ for Helium
\cite{2011Sci...332...69A}. Note, for the fit of proton spectrum
in a wider energy range, as shown in Fig. \ref{fig:proton}, the
spectral difference is only about $0.1$.

\begin{figure*}[!htb]
\includegraphics[width=0.45\textwidth]{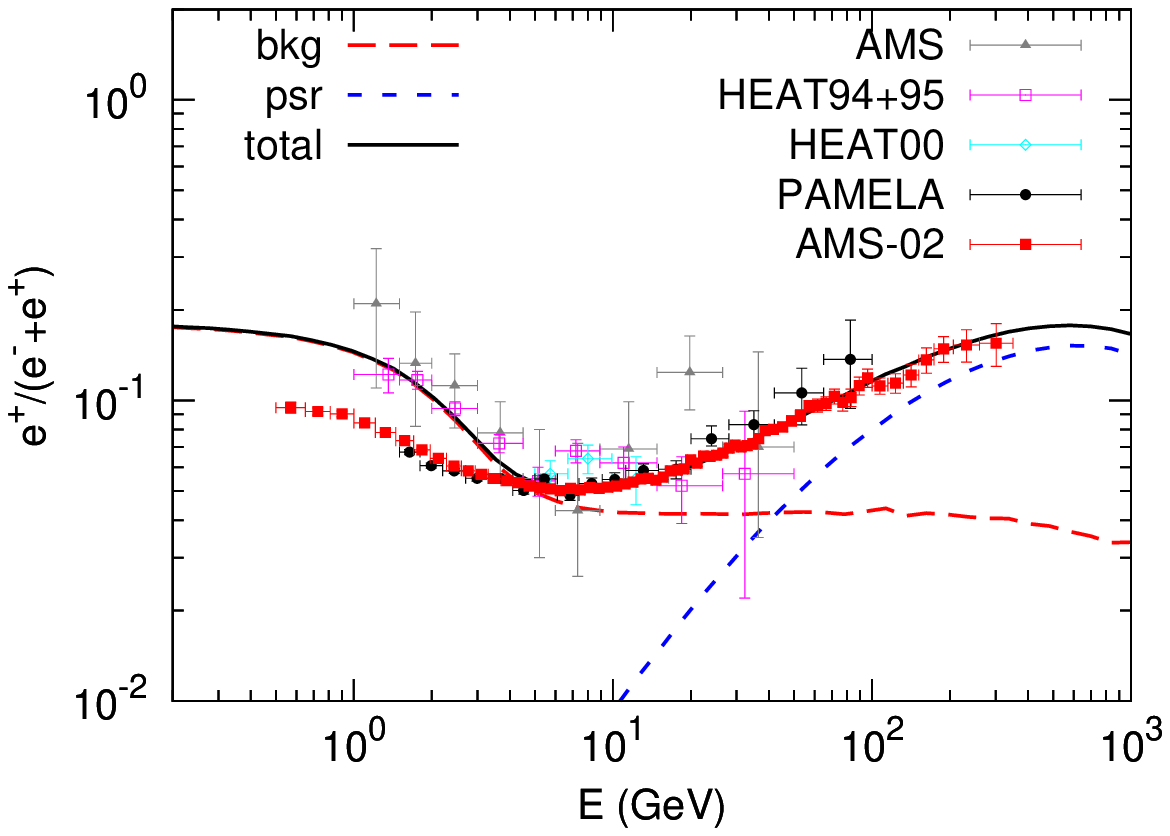}
\includegraphics[width=0.45\textwidth]{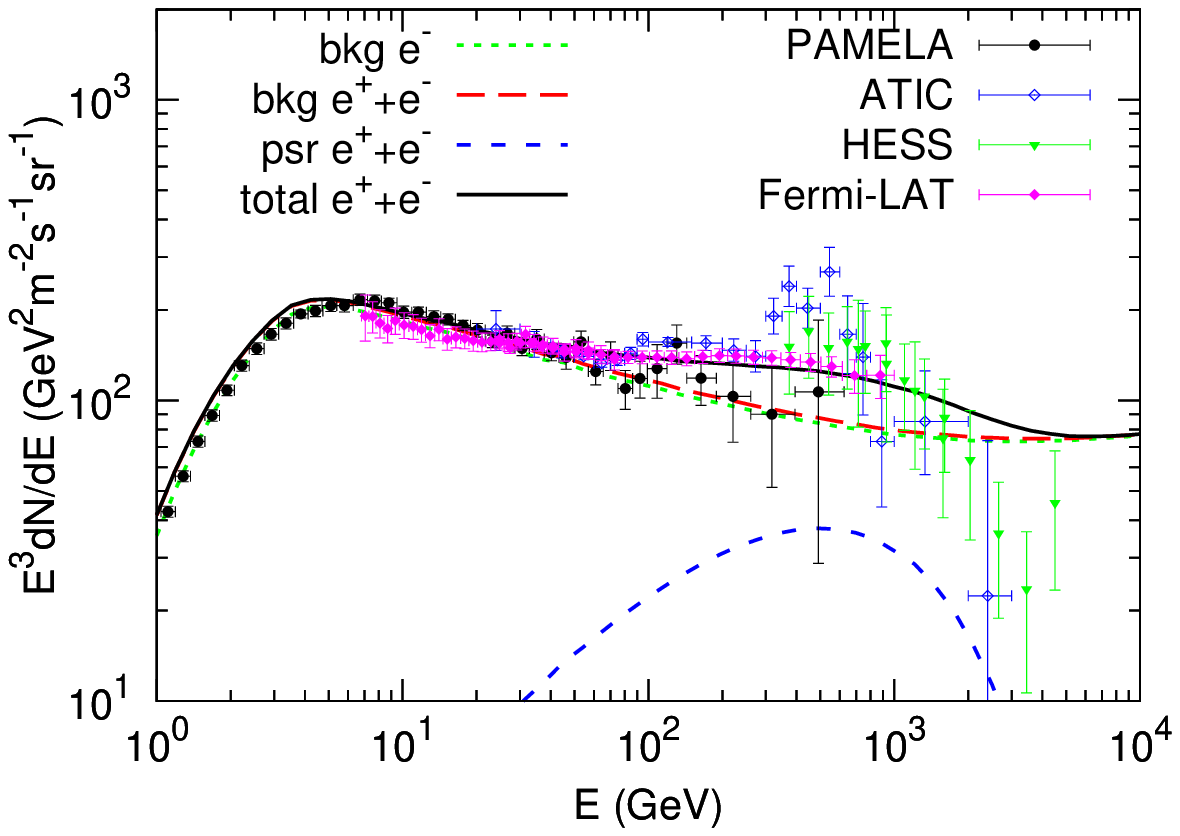}
\caption{Same as Fig. \ref{fig:psr_break} but for a log-parabolic
approximation of the spectral hardening of the primary electron spectrum.
\label{fig:psr_curve}}
\end{figure*}

It is also possible that the spectral hardening is not a break but
a smooth hardening instead, as shown in many models
\cite{2011PhRvD..84d3002Y, 2013ApJ...763...47P}. We may use a
log-parabolic function to approximate the smooth hardening of the
electron spectrum. Fig. \ref{fig:psr_curve} presents the results
of the fit with log-parabolic shape of the primary electron
spectrum. We find the fit is also improved, with the minimum
$\chi^2$ value slightly larger than that with $p_{\rm br,2}^e$
free. The fitting parameters are given in Table \ref{table:psr}.

\begin{figure}[!htb]
\includegraphics[width=0.45\textwidth]{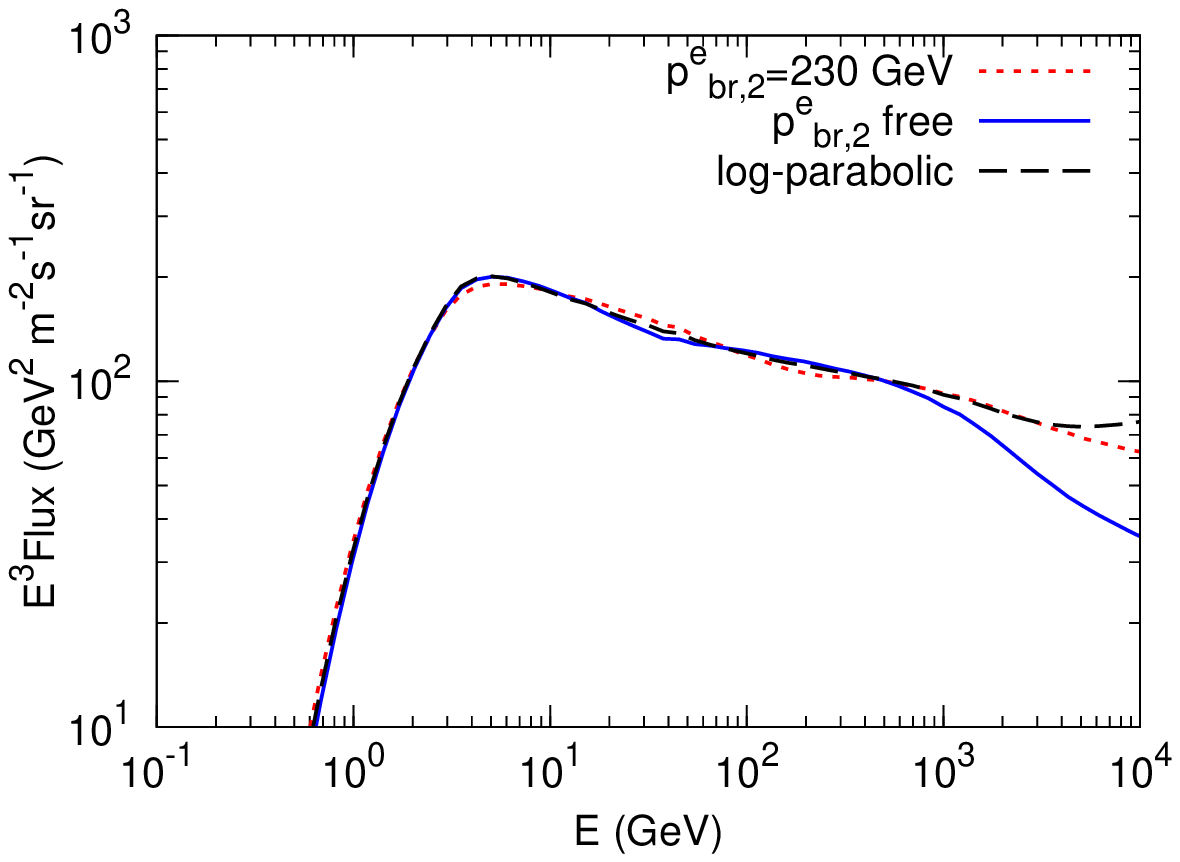}
\caption{Expected total fluxes of the pure electrons for the three
fits corresponding to Figs. \ref{fig:psr_break2} - \ref{fig:psr_curve}.
\label{fig:eflux}}
\end{figure}

From above we see that including a spectral hardening of the primary
electron spectrum, both the PAMELA, AMS-02 and Fermi data can be well
fitted with a single component of the extra sources. It is a natural
expectation that there is a hardening in the primary electron
spectrum, given the observed hardening of the CR nuclei. However,
the position of the break might be different from that of nuclei. It
is a problem needs to be further understood theoretically. If the
hardening of the primary electron spectrum can be confirmed, it would
be important to understand the origin and acceleration of the Galactic
CRs. Since AMS-02 could measure the pure electron spectrum with much
higher precision that PAMELA, we give the expected pure electron
spectra in Fig. \ref{fig:eflux} for the above three cases of the
hardening. The future AMS-02 data may test the existence and detailed
shape of the primary electron spectrum.

\begin{figure*}[!htb]
\includegraphics[width=0.45\textwidth]{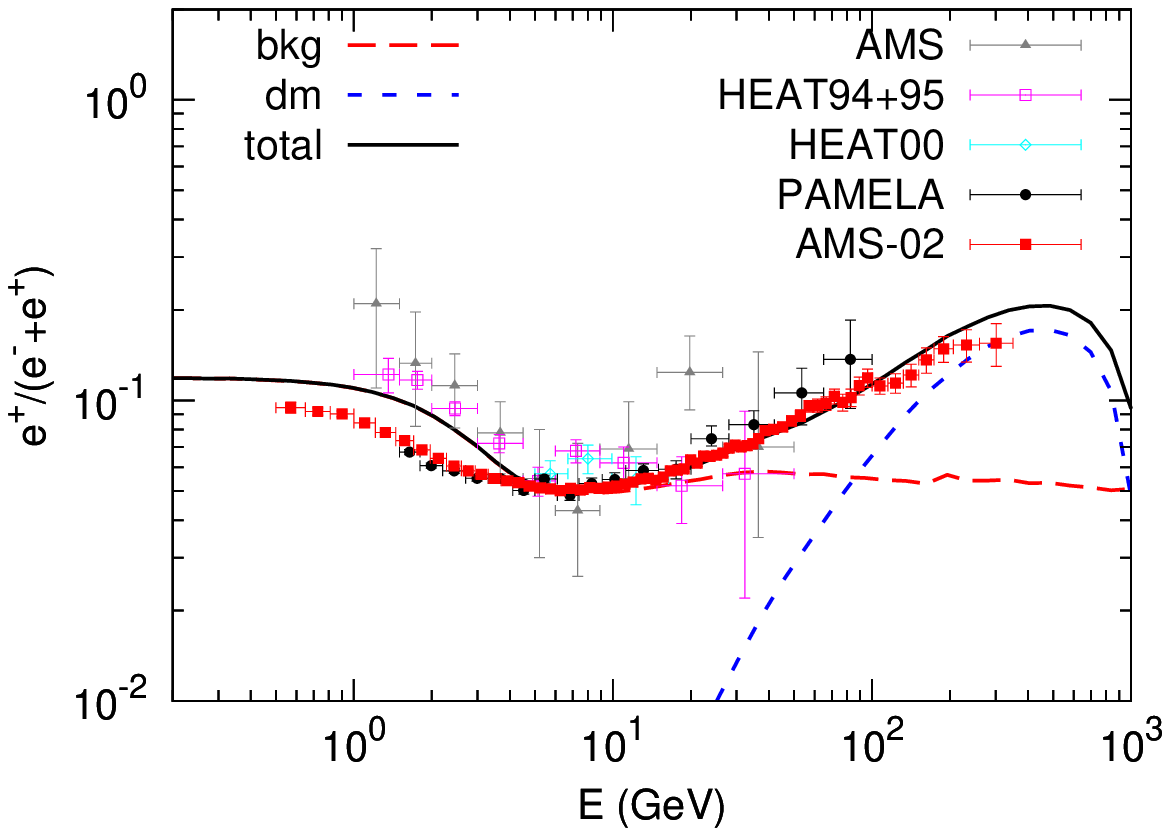}
\includegraphics[width=0.45\textwidth]{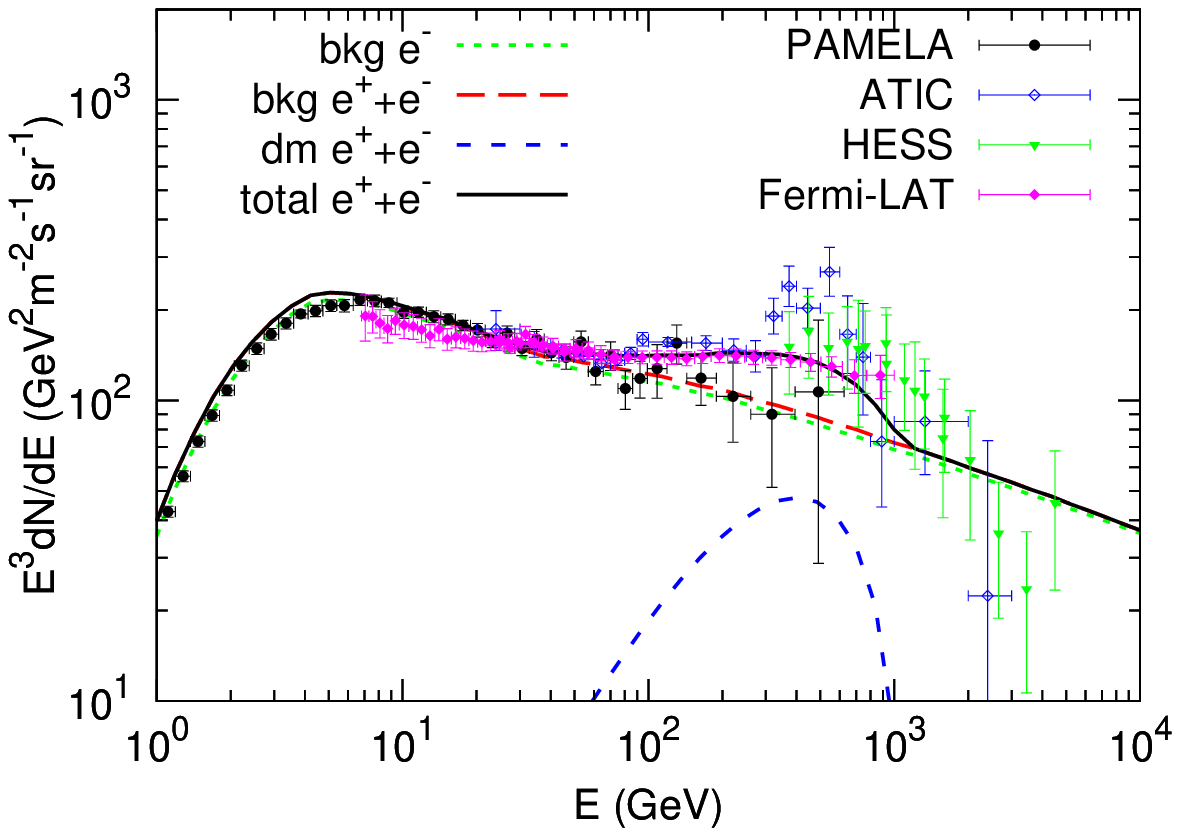}
\caption{Same as Fig. \ref{fig:psr_break} but for the DM annihilation
into a pair of muons as the extra source of the positrons and electron.
\label{fig:mu_break}}
\end{figure*}

\begin{figure*}[!htb]
\includegraphics[width=0.45\textwidth]{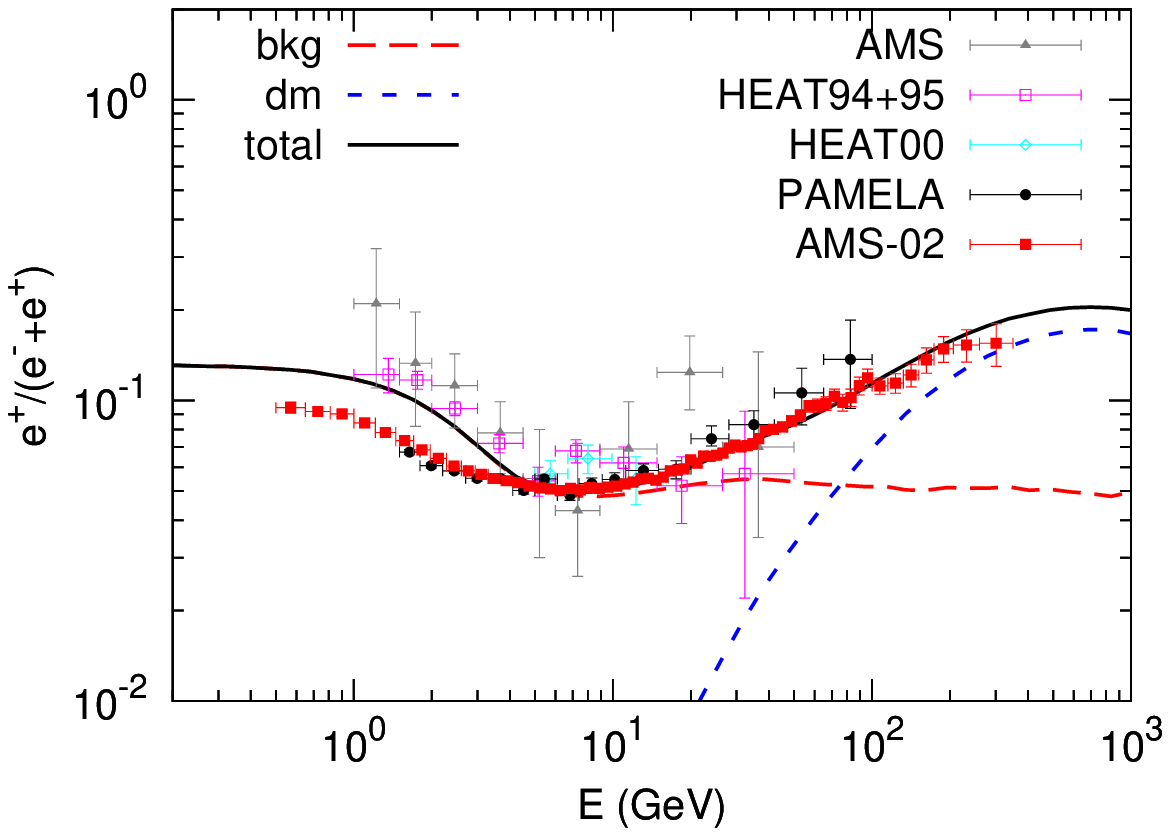}
\includegraphics[width=0.45\textwidth]{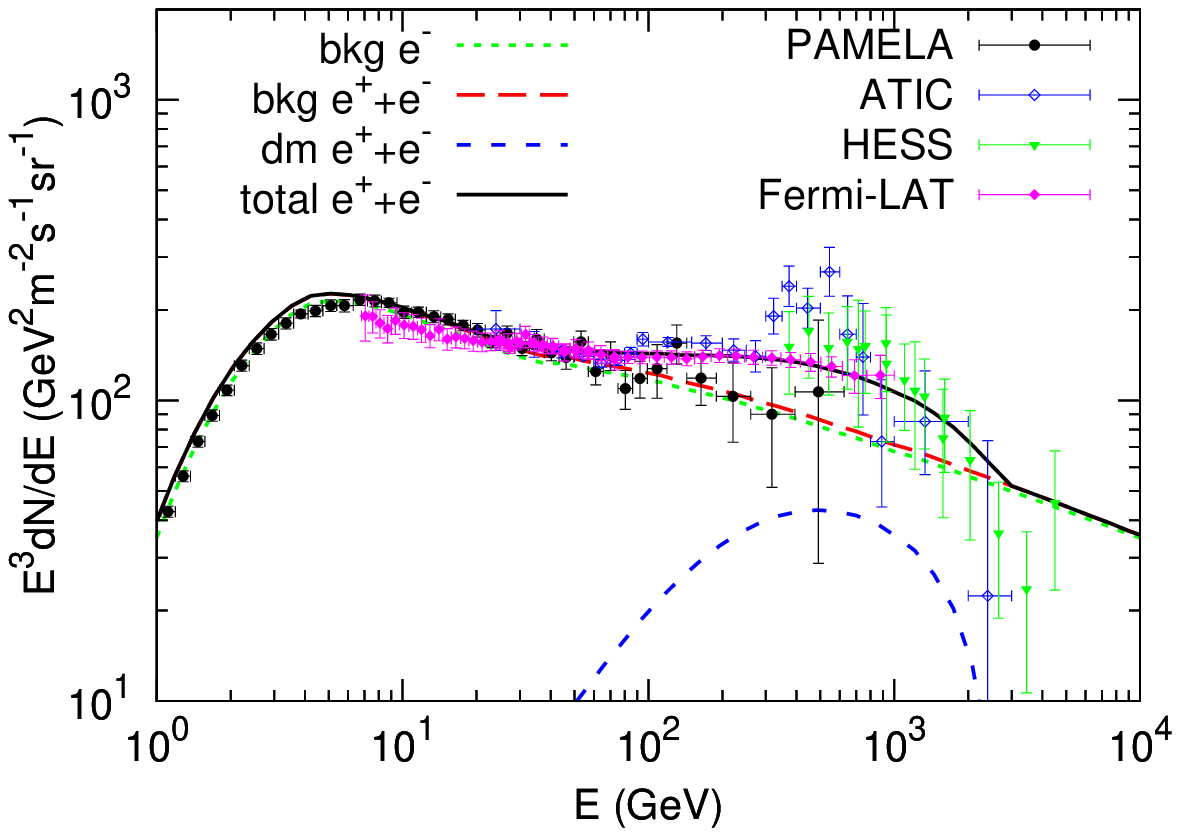}
\caption{Same as Fig. \ref{fig:psr_break} but for the DM annihilation
into a pair of tauons as the extra source of the positrons and electron.
\label{fig:tau_break}}
\end{figure*}

Finally we discuss the DM annihilation as the sources of the $e^{\pm}$.
The annihilation final states are assumed to be $\mu^+\mu^-$ and
$\tau^+\tau^-$. The primary electron spectrum is parameterized with
Eq. (1), and $p_{\rm br,2}^e$ is allowed to be free in the fit. The
fitting results are shown in Figs. \ref{fig:mu_break} and
\ref{fig:tau_break}, for $\mu^+\mu^-$ and $\tau^+\tau^-$ final states
respectively. The fitting parameters are compiled in Table \ref{table:dm}.
It is shown that the DM models can give comparable fittings to the
data compared with pulsars. The break momentum of the primary electrons,
$p_{\rm br,2}^e$ is also similar with that derived in the pulsar
scenario, and is smaller than $p_{\rm br,2}^p\approx230$ GeV.

\begin{figure}[!htb]
\includegraphics[width=0.45\textwidth]{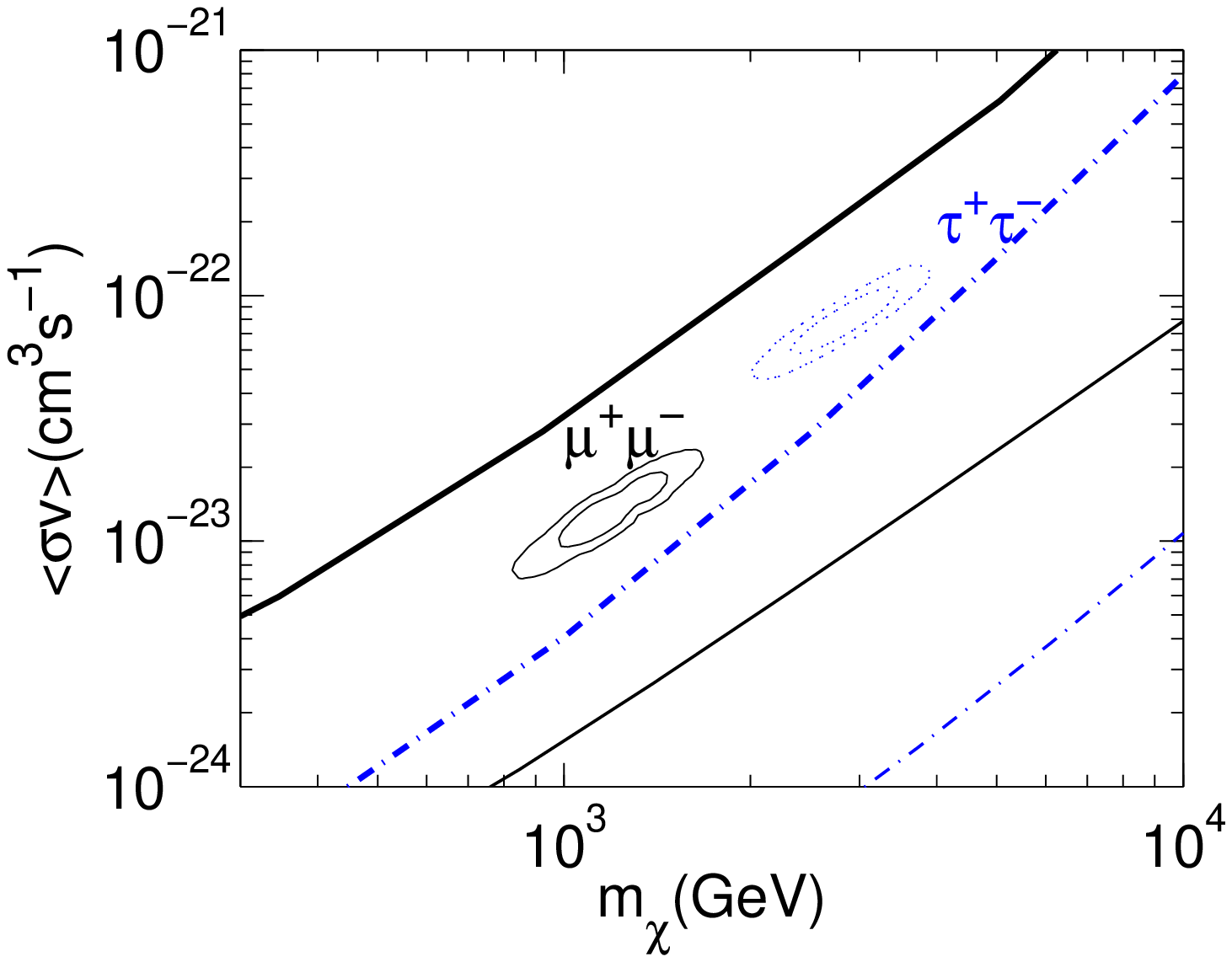}
\caption{$1\sigma$ and $2\sigma$ parameter regions on the
$m_{\chi}-\sv$ plane for the DM annihilation scenario. The lines
show the $95\%$ upper limit of Fermi $\gamma$-ray observations of
the Galactic center (thin lines, with different normalization of
the local density corrected, \cite{2012JCAP...11..048H}) and dwarf
galaxies (thick lines, \cite{Drlica-Wagner2012}) for $\mu^+\mu^-$
(black solid) and $\tau^+\tau^-$ (blue dashed-dotted) channels
respectively.
\label{fig:msv_mt}}
\end{figure}

The $1\sigma$ and $2\sigma$ favored regions on the $m_{\chi}-\sv$
parameter plane are given in Fig. \ref{fig:msv_mt}. For $\mu^+\mu^-$
channel DM with mass $0.8-1.5$ TeV is favored, while for $\tau^+\tau^-$
channel the mass is obtained to be $2-4$ TeV. The boost factor of the
annihilation cross section compared with the natural value to give
the proper relic density is about hundred to thousand. Such results
do not differ much from the ones obtained through fitting the
PAMELA positron fraction and the Fermi/HESS $e^{\pm}$ data
\cite{2010NuPhB.831..178M}.

The exclusion limits on the DM annihilation into $\mu^+\mu^-$ and
$\tau^+\tau^-$ pairs by $\gamma$-rays from the Galactic center
(thin lines, \cite{2012JCAP...11..048H}) and the dwarf galaxies
(thick lines, \cite{Drlica-Wagner2012}) are also plotted in Fig.
\ref{fig:msv_mt}. The results show that for the $\tau^+\tau^-$
channel the Fermi $\gamma$-rays always give very strong
constraints on the annihilation cross section. The constraints for
the $\mu^+\mu^-$ channel is weaker. The Galactic center
$\gamma$-rays tend to exclude the parameter space to explain the
$e^{\pm}$ excesses. However it may suffer from the uncertainties
of the density profile of DM in the halo center. The more robust
limits from the dwarf galaxies can not exclude the favored
parameter region.

\section{Conclusion and discussion}

The study of the highly precise data of positron fraction in CRs
reported by AMS-02, as well as the pure electron spectrum measured
by PAMELA and the total $e^{\pm}$ spectra measured by Fermi and
HESS, shows that it is difficult to use a single component of the
extra sources to explain the $e^{\pm}$ excesses (Paper I,
\cite{2013arXiv1304.1482Y}). In this work we show that an additional 
break of the primary electron spectrum can improve the fit significantly. 
The best fitting break momentum is about $40-50$ GeV and the spectral 
difference $\gamma_2-\gamma_1$ is $\sim0.3-0.4$. As a comparison, the 
break momentum of protons is about $230$ GeV, and the spectral 
difference is $\sim 0.1$. The hardening behavior of the electron 
spectrum is different from that of nuclei, which makes the 
understanding of the fine structures of the CR spectra non-trivial.

The different behaviors of the spectral hardening between nuclei 
and electrons are probably due to the fact that high energy electrons 
should come from nearby regions, and less number of relevant sources 
leads to larger fluctuations of the electron spectrum than that of nuclei.
It is also possible that, if one or several nearby sources are responsible
to the spectral hardening, the accelerated electron-to-proton ratio is
higher for these sources.

In the presence of a hardening of the primary electron spectrum,
both the pulsar and DM scenarios can give comparable fit to the 
data. However, the DM scenario are strongly constrained by the 
$\gamma$-rays, especially for the tauon final state. We would like 
to point out that it will be equivalent to take the harder part of 
the electron spectrum and to drop the constraints from the PAMELA
electron data. In such ways both the AMS-02 positron fraction and 
Fermi total $e^\pm$ spectrum can be fitted simultaneously.

The AMS-02 will measure the electron spectrum with high precision
in the near future. Whether there is a hardening in the electron
spectrum or a lower $e^\pm$ total spectrum than Fermi will soon be
answered by AMS-02.

\begin{table*}[!htb]
\caption {Fitting results of pulsar-like model for different primary electron spectra}
\begin{tabular}{c|cc|cc|cc}
\hline \hline
 & \multicolumn{2}{c|}{$p_{\rm br,2}^e=230$ GeV} & \multicolumn{2}{c|}{$p_{\rm br,2}^e$ free} & \multicolumn{2}{c}{log-parabolic} \\
$\chi^2_{\rm min}/{\rm dof}$ & \multicolumn{2}{c|}{$235.3/150$} & \multicolumn{2}{c|}{$152.5/149$} & \multicolumn{2}{c}{$171.4/150$} \\
\hline
parameters & best & mean & best & mean & best & mean \\
\hline
$\log(A_e\footnotemark[1])$ & $-8.971$ & $-8.973\pm0.005$ & $-8.978$ & $-8.977\pm0.007$ & $-8.980$ & $-8.976\pm0.005$ \\
$\gamma_0$  & $1.504$ & $1.515\pm0.015$ & $1.624$ & $1.606\pm0.054$ & $1.536$ & $1.558\pm0.042$ \\
$\log(p_{\rm br,1}^e/{\rm MeV})$ & $3.587$ & $3.592\pm0.021$ & $3.625$ & $3.622\pm0.030$ & $3.630$ & $3.633\pm0.021$ \\
$\gamma_1$  & $2.683$ & $2.683\pm0.011$ & $2.851$ & $2.839\pm0.023$ & $3.252$ & $3.283\pm0.056$ \\
$\log(p_{\rm br,2}^e/{\rm MeV})$ & --- & --- & $4.630$ & $4.638\pm0.068$ & --- & --- \\
$\gamma_2$  & $2.368$ & $2.370\pm0.031$ & $2.536$ & $2.533\pm0.025$ & $0.101$ & $0.106\pm0.009$ \\
$\log(A_{\rm psr}\footnotemark[2])$ & $-25.036$ & $-25.066\pm0.168$ & $-27.406$ & $-26.981\pm0.397$ & $-26.502$ & $-26.547\pm0.359$ \\
$\alpha$  & $1.881$ & $1.875\pm0.034$ & $1.434$ & $1.516\pm0.077$ & $1.599$ & $1.592\pm0.071$ \\
$\log(p_c/{\rm MeV})$ & $6.242$ & $6.280\pm0.103$ & $6.022$ & $6.074\pm0.087$ & $6.053$ & $6.043\pm0.095$ \\
$c_{e^+}$ & $1.460$ & $1.464\pm0.077$ & $2.244$ & $2.177\pm0.104$ & $1.916$ & $1.972\pm0.103$ \\
$\phi/{\rm MV}$ & $561$ & $558\pm29$ & $735$ & $716\pm37$ & $629$ & $653\pm36$ \\
\hline
\hline
\end{tabular}\vspace{3mm}\\
\footnotemark[1]{Normalization at 25 GeV in unit of
cm$^{-2}$s$^{-1}$sr$^{-1}$MeV$^{-1}$.}\\
\footnotemark[2]{Normalization at 1 MeV in unit of
cm$^{-3}$s$^{-1}$MeV$^{-1}$.}
\label{table:psr}
\end{table*}

\begin{table*}[!htb]
\caption {Fitting results of DM annihilation model}
\begin{tabular}{c|cc|cc}
\hline \hline
 & \multicolumn{2}{c|}{$\mu^+\mu^-$} & \multicolumn{2}{c}{$\tau^+\tau^-$} \\
$\chi^2_{\rm min}/{\rm dof}$ & \multicolumn{2}{c|}{$186.6/150$} & \multicolumn{2}{c|}{$160.5/150$} \\
\hline
parameters & best & mean & best & mean \\
\hline
$\log(A_e\footnotemark[1])$ & $-8.970$ & $-8.970\pm0.007$ & $-8.979$ & $-8.978\pm0.008$ \\
$\gamma_0$  & $1.766$ & $1.768\pm0.039$ & $1.718$ & $1.712\pm0.046$ \\
$\log(p_{\rm br,1}^e/{\rm MeV})$ & $3.690$ & $3.688\pm0.026$ & $3.678$ & $3.671\pm0.027$ \\
$\gamma_1$  & $2.938$ & $2.943\pm0.012$ & $2.925$ & $2.921\pm0.014$ \\
$\log(p_{\rm br,2}^e/{\rm MeV})$ & $4.611$ & $4.597\pm0.041$ & $4.554$ & $4.563\pm0.048$ \\
$\gamma_2$  & $2.541$ & $2.547\pm0.023$ & $2.547$ & $2.548\pm0.019$ \\
$\log(m_\chi/{\rm GeV})$  & $3.063$ & $3.071\pm0.061$ & $3.486$ & $3.451\pm0.056$ \\
$\log(\sv/{\rm cm^3s^{-1}})$ & $-22.897$ & $-22.885\pm0.105$ & $-22.043$ & $-22.104\pm0.090$ \\
$c_{e^+}$ & $2.683$ & $2.687\pm0.059$ & $2.524$ & $2.522\pm0.067$ \\
$\phi/{\rm MV}$ & $839$ & $849\pm30$ & $796$ & $796\pm32$ \\
\hline
\hline
\end{tabular}\vspace{3mm}\\
\footnotemark[1]{Normalization at 25 GeV in unit of
cm$^{-2}$s$^{-1}$sr$^{-1}$MeV$^{-1}$.}
\label{table:dm}
\end{table*}

\appendix

\section{Results of the background positrons and electrons}

For the convenience of use we tabulate the fluxes of the background
positrons and electrons calculated with the best fitting parameters
of the pulsar models in Table \ref{table:bkg}. For DM models the
results have little difference. Note for background positrons an
additional factor $c_{e^+}$ as given in Table \ref{table:psr} needs
to be multiplied. Here the local interstellar fluxes are given.
If one wants to better reproduce the low energy electron spectrum,
the solar modulation with modulation potential given in Table
\ref{table:psr} is necessary.

\begin{table*}[!htb]
\caption {Local interstellar fluxes of the background positrons and electrons}
\begin{tabular}{c|c|ccc}
\hline\hline
$E$(GeV) & \multicolumn{4}{c}{$E^2F$(GeV m$^{-2}$ s$^{-1}$ sr$^{-1}$)} \\
\hline
         & $e^+$ & $e^-$($p_{\rm br,2}^e=230$ GeV) & $e^-$($p_{\rm br,2}^e$ free) & $e^-$(log-parabolic) \\
\hline
$1.0000E-01$ & $1.0085E+01$ & $6.3746E+01$ & $1.2352E+02$ & $8.3203E+01$ \\
$1.2649E-01$ & $1.3343E+01$ & $8.1775E+01$ & $1.5669E+02$ & $1.0641E+02$ \\
$1.5999E-01$ & $1.6996E+01$ & $1.0265E+02$ & $1.9495E+02$ & $1.3312E+02$ \\
$2.0236E-01$ & $2.0939E+01$ & $1.2667E+02$ & $2.3747E+02$ & $1.6386E+02$ \\
$2.5595E-01$ & $2.4880E+01$ & $1.5278E+02$ & $2.8359E+02$ & $1.9708E+02$ \\
$3.2375E-01$ & $2.8681E+01$ & $1.8086E+02$ & $3.3104E+02$ & $2.3162E+02$ \\
$4.0949E-01$ & $3.2094E+01$ & $2.0769E+02$ & $3.7547E+02$ & $2.6486E+02$ \\
$5.1795E-01$ & $3.4604E+01$ & $2.3058E+02$ & $4.1054E+02$ & $2.9230E+02$ \\
$6.5513E-01$ & $3.5781E+01$ & $2.4683E+02$ & $4.3241E+02$ & $3.1091E+02$ \\
$8.2864E-01$ & $3.5182E+01$ & $2.5437E+02$ & $4.3443E+02$ & $3.1729E+02$ \\
$1.0481E+00$ & $3.2114E+01$ & $2.4778E+02$ & $4.1198E+02$ & $3.0636E+02$ \\
$1.3257E+00$ & $2.7285E+01$ & $2.3049E+02$ & $3.7033E+02$ & $2.8109E+02$ \\
$1.6768E+00$ & $2.1238E+01$ & $2.0411E+02$ & $3.1604E+02$ & $2.4608E+02$ \\
$2.1210E+00$ & $1.4943E+01$ & $1.7304E+02$ & $2.5584E+02$ & $2.0548E+02$ \\
$2.6827E+00$ & $9.6236E+00$ & $1.4278E+02$ & $2.0096E+02$ & $1.6662E+02$ \\
$3.3932E+00$ & $5.7713E+00$ & $1.1408E+02$ & $1.5456E+02$ & $1.3231E+02$ \\
$4.2919E+00$ & $3.3159E+00$ & $8.6156E+01$ & $1.1499E+02$ & $1.0119E+02$ \\
$5.4287E+00$ & $1.9367E+00$ & $6.0846E+01$ & $7.8414E+01$ & $7.0057E+01$ \\
$6.8665E+00$ & $1.1805E+00$ & $4.2413E+01$ & $5.2323E+01$ & $4.7256E+01$ \\
$8.6851E+00$ & $7.5988E-01$ & $2.9862E+01$ & $3.5204E+01$ & $3.2186E+01$ \\
$1.0985E+01$ & $5.1230E-01$ & $2.1185E+01$ & $2.3941E+01$ & $2.2173E+01$ \\
$1.3895E+01$ & $3.5859E-01$ & $1.5280E+01$ & $1.6600E+01$ & $1.5669E+01$ \\
$1.7575E+01$ & $2.5470E-01$ & $1.1079E+01$ & $1.1644E+01$ & $1.1211E+01$ \\
$2.2230E+01$ & $1.8315E-01$ & $8.0717E+00$ & $8.1879E+00$ & $8.0348E+00$ \\
$2.8118E+01$ & $1.3379E-01$ & $5.9131E+00$ & $5.8546E+00$ & $5.8577E+00$ \\
$3.5565E+01$ & $9.7597E-02$ & $4.3215E+00$ & $4.2067E+00$ & $4.2699E+00$ \\
$4.4984E+01$ & $7.3397E-02$ & $3.2332E+00$ & $3.1587E+00$ & $3.2056E+00$ \\
$5.6899E+01$ & $5.3549E-02$ & $2.3330E+00$ & $2.3430E+00$ & $2.3364E+00$ \\
$7.1969E+01$ & $3.9877E-02$ & $1.7106E+00$ & $1.7699E+00$ & $1.7348E+00$ \\
$9.1030E+01$ & $2.9733E-02$ & $1.2613E+00$ & $1.3334E+00$ & $1.2936E+00$ \\
$1.1514E+02$ & $2.2265E-02$ & $9.3209E-01$ & $1.0104E+00$ & $9.6957E-01$ \\
$1.4563E+02$ & $1.6558E-02$ & $6.8757E-01$ & $7.5661E-01$ & $7.2384E-01$ \\
$1.8421E+02$ & $1.2349E-02$ & $5.1494E-01$ & $5.6947E-01$ & $5.4575E-01$ \\
$2.3300E+02$ & $9.2290E-03$ & $3.9036E-01$ & $4.2686E-01$ & $4.1269E-01$ \\
$2.9471E+02$ & $6.8405E-03$ & $3.0270E-01$ & $3.1791E-01$ & $3.1194E-01$ \\
$3.7276E+02$ & $5.0884E-03$ & $2.3532E-01$ & $2.3806E-01$ & $2.3722E-01$ \\
$4.7149E+02$ & $3.7641E-03$ & $1.8276E-01$ & $1.7733E-01$ & $1.8072E-01$ \\
$5.9636E+02$ & $2.7641E-03$ & $1.4101E-01$ & $1.3197E-01$ & $1.3743E-01$ \\
$7.5431E+02$ & $2.0308E-03$ & $1.0876E-01$ & $9.7899E-02$ & $1.0576E-01$ \\
$9.5410E+02$ & $1.4738E-03$ & $8.3842E-02$ & $7.2376E-02$ & $8.1243E-02$ \\
$1.2068E+03$ & $1.0837E-03$ & $6.5258E-02$ & $5.4198E-02$ & $6.3344E-02$ \\
$1.5264E+03$ & $7.7609E-04$ & $5.0293E-02$ & $4.0077E-02$ & $4.9167E-02$ \\
$1.9307E+03$ & $5.5115E-04$ & $3.8733E-02$ & $2.9646E-02$ & $3.8295E-02$ \\
$2.4421E+03$ & $3.8831E-04$ & $2.9902E-02$ & $2.1981E-02$ & $3.0025E-02$ \\
$3.0888E+03$ & $2.7001E-04$ & $2.3065E-02$ & $1.6307E-02$ & $2.3726E-02$ \\
$3.9069E+03$ & $1.8373E-04$ & $1.7810E-02$ & $1.2022E-02$ & $1.8746E-02$ \\
$4.9417E+03$ & $1.2265E-04$ & $1.3675E-02$ & $8.8979E-03$ & $1.4890E-02$ \\
$6.2506E+03$ & $8.0038E-05$ & $1.0560E-02$ & $6.5895E-03$ & $1.1899E-02$ \\
$7.9060E+03$ & $5.0413E-05$ & $8.1294E-03$ & $4.8656E-03$ & $9.5347E-03$ \\
$1.0000E+04$ & $3.0695E-05$ & $6.2648E-03$ & $3.5951E-03$ & $7.6744E-03$ \\
\hline
\hline
\end{tabular}
\label{table:bkg}
\end{table*}

\acknowledgments
This work is supported by 973 Program under Grant No. 2013CB837000 
and the National Natural Science Foundation of China under Grant 
Nos. 11075169, 11105155.

\bibliographystyle{apsrev}
%\bibliography{refs}
\bibliography{/home/yuanq/work/cygnus/tex/refs}

\end{document}